\documentclass[prl,aps,showpacs,twocolumn,superscriptaddress]{revtex4-1}
\usepackage{amsmath,amssymb,graphicx}
\usepackage{amsmath}
\usepackage{ulem}
\usepackage{enumitem}

\begin{document}

\title{Generation of ultra broadband coherent supercontinuum in tapered and dispersion managed silicon nanophotonic waveguides}

\author{Charles Ciret}
\email{charles.ciret@ulb.ac.be}
\author{Simon-Pierre Gorza}
\affiliation{OPERA-Photonique, Universit\'e libre de Bruxelles (ULB), 50 av. F.D. Roosevelt, CP194/5, B-1050 Bruxelles, Belgium}




\begin{abstract}
Tapered and dispersion managed (DM) silicon nanophotonic waveguides are investigated for the generation of optimal ultra broadband supercontinuum (SC). DM waveguides are structures showing a longitudinally dependent group velocity dispersion that results from the variation of the waveguide width with the propagation distance. For the generation of optimal SC, a genetic algorithm has been used to find the best dispersion map. This allows for the generation of highly coherent supercontinuums that span over 1.14 octaves from 1300~nm to 2860~nm and 1.25 octaves from 1200~nm to 2870~nm at -20~dB level for the tapered and DM waveguides respectively, for a 2\,$\mu$m, 200 fs and 6.4 pJ input pulse. The comparison of these two structures with the usually considered optimal fixed width waveguide shows that the SC is broader and flatter in the more elaborated DM waveguide, while the high coherence is ensured by the varying dispersion.
\end{abstract}


\maketitle

\section{Introduction}

Supercontinuum generation (SCG), resulting from the dramatic spectral broadening of an input narrowband pulse during its propagation in a nonlinear waveguide, has been extensively studied by the scientific community since its discovery in the early 1970s \cite{Alfano:70,Dudley:06}. Since then, numerous applications of these wide spectra have emerged in frequency metrology \cite{Jones:00}, telecommunications \cite{Smirnov:06} or in spectroscopy \cite{Hartl:01}, promoting even more their interests. In particular, the advent of photonic crystal fibers (PCFs) has led to the generation of supercontinuum on more than three octaves \cite{Jiang:15,Belli:15} thanks to their design flexibility and their tight light confinement capabilities. Due to the crucial role played by the waveguide dispersion properties on the SCG dynamics, the transverse structure of PCFs have been engineered, using for instance genetic algorithms, and this has led to the generation of quasi on-demand output spectra \cite{Zhang:09, Arteaga-Sierra:14}. Other studies have investigated the possibility of engineering the group velocity dispersion (GVD) along the length of the waveguide in order to reach targeted output spectra\,\citep{Bendahmane:13}. In the framework of optical fiber, we can cite studies performed in simple structures such as tapered single mode fibers \cite{Lu:04} or in stepwise dispersion decreasing optical fibers \cite{Abrardi:07}.  
SCG in integrated structures has also raised a lot of attention, and demonstrations of on-chip SCG in chalcogenide \cite{Lamont:08}, silicon \cite{Leo:14}, amorphous silicon \cite{Leo:14b}, silicon nitride \cite{Johnson:15} or in III-V materials \cite{Dave:15} have been reported. Beside their millimeter size, these waveguides possess a nonlinear coefficient several order of magnitude larger than in PCF, making them the ideal candidates for potential low power integrated applications. 
In these nanophotonic structures, the waveguide dispersion is mainly determined by the waveguide geometry.  
Engineered dispersion waveguide thus means that this geometry, i.e. often only the waveguide width due to fabrication constraints, has been optimized to allow for the broadest SC given the input pulse and the material properties. Most of the previous studies have thus only considered fixed width (FW) structures in which the dispersion properties are maintained along the propagation. In~\cite{Hu:13, Hu:15} the authors investigate glass on silica taper waveguides, in which the dispersion properties are modified by varying the waveguide height. However the fabrication process of integrated platforms, such as semiconductor on insulator, allow for the realization of more elaborated structures than FW and tapered waveguides, without any increase of the fabrication complexity, contrary to dispersion varying glass based fibers and planar rib waveguides. One can thus make advantage of this extra degree of freedom to optimize the desired properties of the SC.

In this paper, we focus on the generation of broad and flat coherent supercontinuum by managing the GVD properties through an adjustment of the waveguide width along the pulse propagation distance. Surprisingly, even though the crucial role played by the dispersion on the SCG is well known, single taper structures have only been considered in the context of glass-based integrated structures. In optical fibers, more complex structures have for instance been studied for the control of the spectral position of Raman soliton \cite{Bendahmane:13}. Refereeing to the dispersion maps encountered in optical telecommunication, these types of structures are called dispersion managed (DM) waveguides. We consider in this work SCG in silicon on silica nanophotonic waveguides, as it is an important and well established platform. SCG in tapered silicon structures will first be investigated since they have been recognized to improve the SC bandwidth in glass-based integrated waveguides and fibers. More elaborated dispersion maps will then be considered. Note however, that tapers can be viewed as a particular case of DM waveguides.  The SCG in the DM photonic waveguides is numerically simulated by using the generalized nonlinear Schrödinger equation coupled to the free-carriers evolution equation that applies for silicon nano-waveguides below the half bandgap [see Eq.\,(\ref{EqGNLS}-\ref{FC})]. In silicon waveguides, the nonlinear dynamics is governed by the Kerr nonlinear effect, the two-photon absorption (TPA) and the high-order dispersion while, contrary to SCG in fibers, the Raman effect does not play a significant role. Similarly to\,\cite{Bendahmane:13}, we resorted to a genetic algorithm to find the optimal dispersion map.

\section{Waveguide optimization}

\begin{figure}[t!]
\centering\includegraphics[width=\linewidth]{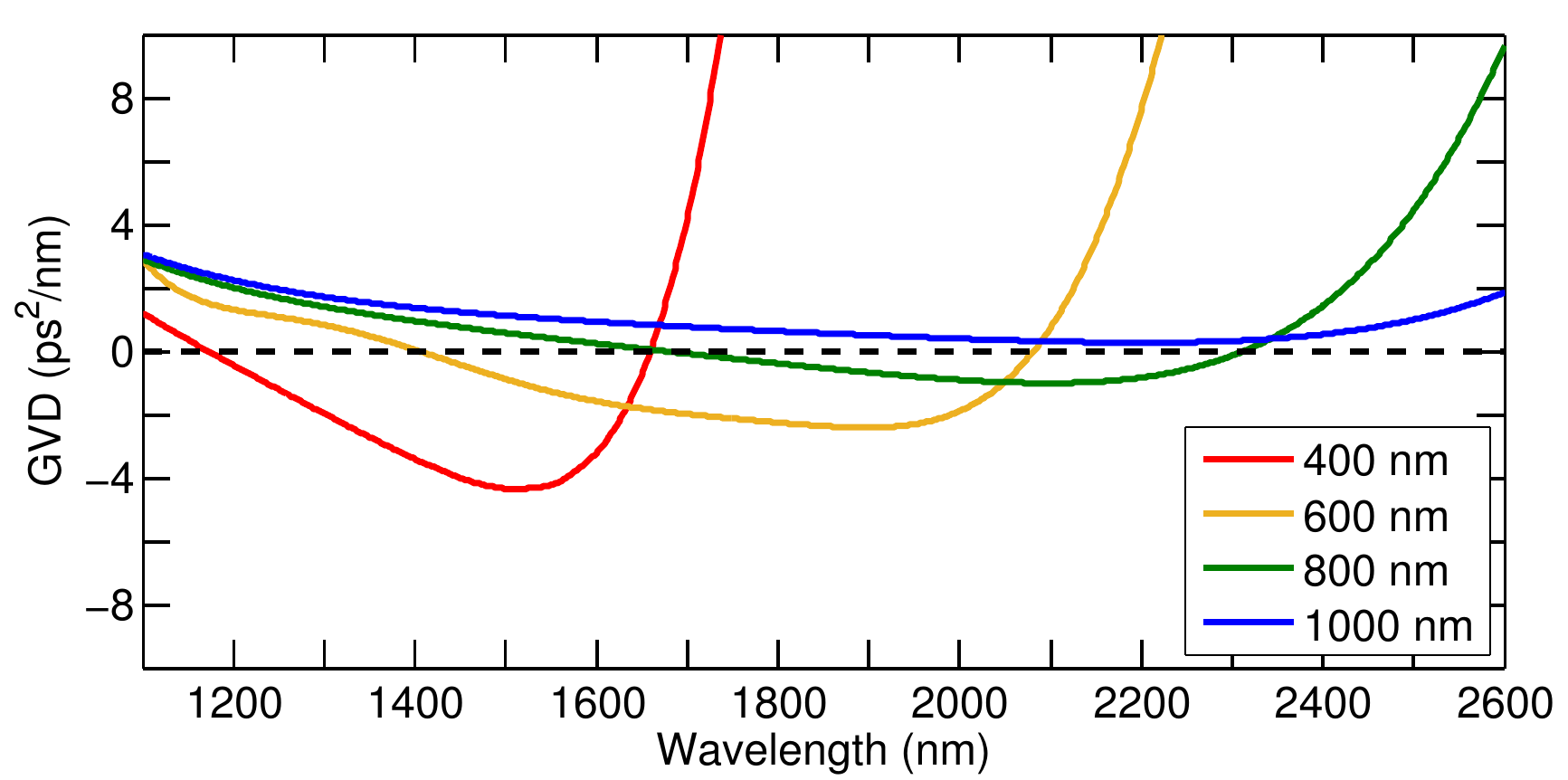}
\caption{(Color online) Group velocity dispersion (GVD) curves for a 220 nm-thick silicon waveguide with different width.}
\label{GVD}
\end{figure}

The underlying dynamics of SCG in nanophotonic integrated silicon structures is now well understood\,\cite{Yin:07,Lin:07}. An intense input pulse, propagating in the anomalous dispersion regime, leads to the formation of a high order soliton, which is temporally compressed in the first steps of the propagation. This is quickly followed by the TPA induced soliton fission and the emission of dispersive waves (DWs), also called Cherenkov radiations. The emission of DWs is essential for generating broad SC in silicon nanophotonic waveguides. Indeed, the Raman induced self frequency shift does not occur as in optical glass fibers because of the small spectral width of the Raman gain in crystalline silicon. DW emission by solitons is a resonant process that couples wavelength in the normal dispersion regime to the soliton wavelength. The exact DW wavelength is thus set by a phase-matching condition which depends on the dispersion properties of the nonlinear waveguide and on the soliton wavelength\,\cite{Akhmediev:95}. As depicted in Fig.~\ref{GVD}, the GVD of silicon waveguides relies strongly on the width of the waveguide. The positions of the two zero dispersion wavelengths (ZDWs), which are crucial for the emission of DWs, are shifted towards longer wavelengths for larger waveguides. Broad SC are thus obtained in structures such that the pump wavelength is located between the two ZDWs in order to excite the DW on each side of the spectrum. As a consequence, the spectra at the output of FW waveguides show usually a spectral density decrease between the pump wavelength and the two DWs (see in Fig.\,\ref{FW} as well as in\,\citep{Leo:14, Johnson:15, Dave:15} for experimental results). In the case of dispersion managed waveguides, we go a step further. By adjusting the width of the waveguide, and therefore the dispersion properties during the propagation, DWs in different spectral domains are expected to be generated, resulting in spectra showing less amplitude variations and, potentially, larger bandwidths\,\cite{Hu:15}. The design of the structure, i.e. the width profile $w(z)$ where $w$ is the waveguide width at position $z$, is optimized using the genetic algorithm named random mutation hill climbing method (RMHC)\cite{Mitchell:96}. The optimization principle is based on the maximization of a particular function, called \textit{fitness function}. As we are seeking for the generation of broad and flat supercontinuums, the fitness function $f$ of the RMHC algorithm has been chosen as:

\begin{align}
f=\frac{1}{\omega_p}&\left(\frac{\displaystyle\int_{\omega_{min}}^{\omega_{max}}\left(\omega-\bar{\omega}\right)^2\left|A(\omega)\right|^2 d\omega}{\displaystyle\int\left|A(\omega)\right|^2d\omega}\right)^{1/2}\nonumber\\&+0.05\left(\mathrm{Var}\left[\log\left(\left|\tilde{A}(\omega)\right|^2\right)\right]\right)^{-1/2},
\label{fitness}
\end{align}
where $\left|A(\omega)\right|^2$ is the output spectral density power at $\omega$ and $\tilde{A}=A/$(1W/Hz).
The first term in Eq.~(\ref{fitness}) is the root mean square spectral width of the generated supercontinuum, computed on the spectral width of interest $[\omega_{min}-\omega_{max}]$. $\bar{\omega}=\displaystyle\int_{\omega_{min}}^{\omega_{max}}\omega\left|A(\omega)\right|^2d\omega/\int\left|A(\omega)\right|^2d\omega$ is the frequency barycenter of the supercontinuum on the optimized frequency range, and $\omega_p$ is the frequency of the input pump pulse. This term thus promotes the emergence of spectral components far from the SC barycenter. In this study we have considered a frequency interval of interest $[\omega_{min}-\omega_{max}]$ corresponding to the wavelength span of [1200~nm$-$2800~nm]. This span corresponds to photon energies below the bandgap ($\lambda\approx1100$\,nm) and to wavelengths that are well guided in the considered 220\,nm-thick waveguides. The second term in Eq.~(\ref{fitness}) is the inverse of the standard deviation of the amplitude of the generated spectrum, in logarithmic scale, where $\mathrm{Var}\left[\log\left(\left|\tilde{A}(\omega)\right|^2\right)\right]=\displaystyle\int_{\omega_{min}}^{\omega_{max}}\left[\log\left(\left|\tilde{A}(\omega)\right|^2\right)-\overline{\log\left(\left|\tilde{A}(\omega)\right|^2\right)}\right]^2 d\omega$, which denotes the spectral variation of the SC generated on the frequency range of interest. $\overline{\log\left(\left|\tilde{A}(\omega)\right|^2\right)}$ is the mean value of the normalized spectral density power in logarithm scale, in the optimized bandwidth. We consider the amplitude variation of the generated spectrum in the fitness function to foster the generation of flat SC. Moreover, this amplitude is considered in logarithmic scale as the spectral components generated far from the pump have an amplitude at least an order of magnitude lower than the spectral amplitude around the input wavelength. Considering logarithmic spectral variation thus enables to reduce the SC power variation without preventing the emergence of broad spectra. The 0.05 coefficient in front of the second term ensures that the two terms in the fitness function have a comparable weight.

In the present study, we consider a 3 mm-long silicon nanophotonic waveguide with a standard height of 220 nm. The optimization of the waveguide width is realized at seven different points located at 0, 0.25, 0.5, 0.75, 1, 2, and 3 mm respectively. Previous demonstrations of SCG in silicon integrated nanophotonic waveguides have shown that most of the SC dynamics results from the self-phase modulation, the temporal compression and the soliton fission which occur within the first millimeters of the propagation (see for instance \cite{Leo:14,Leo:14b,Dave:15,Kuyken:15}). More optimization points have thus been considered at the beginning of the waveguide. We impose a waveguide width between 300 and 1000\,nm to conserve good guidance properties. A linear variation of the width is assumed between the optimization points, the overal structure thus consists of six tapered sections. We verify that the width variations do not exceed 8\,nm/$\mu$m to ensure an adiabatic adaptation of the mode. The optimization of the fitness function by the RMHC method is realized in four steps: 
\begin{enumerate}[label=(\alph*)]
\item  A map containing the values of the width at the seven optimization points is randomly generated. 
\item  The pulse propagation is simulated by numerical integration of the generalized nonlinear Schrödinger equation coupled to the equation describing the free-carriers evolution (see Eq.\,\ref{EqGNLS}-\ref{FC}). The fitness function~(\ref{fitness}) is then computed and the map is stored as a parent map. 
\item  One of the seven points is randomly chosen in the parent map and the corresponding width is changed randomly. The new map is called child map. 
\item  The propagation is again simulated in this new structure by solving the equations ~(\ref{EqGNLS}-\ref{FC}). The value of the fitness function~(\ref{fitness}) is then compared to the value obtained with the previous map. If the fitness of the child map is better than the fitness of the parent map, a better design has been found. The child map becomes the parent map and the step (c) is reiterated. If the fitness of the child map is worse than the fitness of the parent map, the child map is not stored and the step (c) is reiterated from the previous parent map. 
\end{enumerate}
The algorithm stops after 750 iterations and this algorithm is run several times to explore different regions in the parameter space.

\section{Propagation model}

The pulse propagation has been simulated by numerical integration of the Eqs.\,(\ref{EqGNLS}-\ref{FC}) by the split-step Fourier (SSF) algorithm. Theses equations include the specific nonlinear phenomena encountered in nanophotonic silicon structures (free-carriers dispersion (FCD) and absorption (FCA), two-photon absorption (TPA), narrow Raman gain spectrum). The GNLSE Eq.\,(\ref{EqGNLS}) coupled to the free-carriers evolution Eq.\,(\ref{FC}) has been successfully used to simulate supercontinuum generation in nanophotonic waveguides up to 4000 nm broad \cite{Singh:15}, and show very good agreement with experimental results in crystalline silicon \cite{,Leo:14,Leo:15,Ciret:16a}. The two coupled equations read:

\begin{subequations}
\begin{align}
\frac{\partial A(z,t)}{\partial z}&=i\sum_{k=2}^\infty\frac{i^k}{k!}\beta_k\frac{\partial^kA(z,t)}{\partial t^k}-\frac{\alpha_0}{2}A(z,t)-\frac{\alpha_c}{2}(1+i\mu)A(z,t)\nonumber\\ &+i\gamma(1+i\tau_\mathrm{shock}\frac{\partial}{\partial t})A(z,t)\int_{-\infty}^{+\infty} R(t')\left| A(z,t-t')\right|^2 dt'
\label{EqGNLS}
\end{align}
\begin{equation} 
\frac{\partial N_c(z,t)}{\partial t}=\frac{\Im m{(\gamma)}}{\hbar \omega_{p}a_{eff}}|A(z,t)|^4+\frac{\sigma_{AI}}{U_{i} a}N_c(z,t)|A(z,t)|^2-\frac{N_c(z,t)}{\tau_c}.
\label{FC}
\end{equation}
\end{subequations}

In the GNLSE (\ref{EqGNLS}), $\beta_k=\partial^k\beta/\partial\omega^k$ are the dispersion coefficients associated with the Taylor series expansion of the propagation constant $\beta(\omega)$ around the frequency of the input pump. However as the dispersion is computed in the spectral domain in the SSF algorithm, the exact function $\beta(\omega)$ is used without approximation \cite{Dudley:06}. This function has been calculated on the whole wavelength range of interest by finite time difference simulations using the Fimmwave software. $\alpha_0=2 \text{\,dB}/\text{cm}$ is the linear loss coefficient. $\gamma=2\pi n_2/(\lambda_pa_{eff})+i\beta_{TPA}/(2a_{eff})$, is the complex nonlinear parameter, where $n_2$ and $\beta_{TPA}$ are the Kerr and the two-photon absorption coefficients respectively, $a_{eff}$ is the effective mode area and $\lambda_p$ is the input pulse wavelength. From literature \cite{Bristow:07}, $\gamma=$(220+i18)~W$^{-1}$m$^{-1}$ for a 800 nm-wide reference waveguide. Its value is adjusted in function of the waveguide width, through the evolution of the effective mode area computed using the mode solver. As previously shown in \cite{Dudley:06,Singh:15}, the dispersion of the nonlinearity can be modeled by a time derivative term in the nonlinear terms, where $\tau_\mathrm{shock}=1/\omega_{p} - 1/n_2\partial n_2/\partial\omega-1/a_{eff}\partial a_{eff}/\partial\omega$. Taking into account the calculations from the mode solver and the data from literature \cite{Bristow:07}, $1/n_2\partial n_2/\partial\omega=1.7\times10^{-15}$ s$^{-1}$ and $1/a_{eff}\partial a_{eff}/\partial\omega=-1.27\times10^{-15}$ s$^{-1}$ at the input pump wavelength, for the 800\,nm-wide reference waveguide. $R(t')$ is the response function accounting for the instantaneous and delayed Raman contributions to the nonlinearity \cite{Lin:07}.  $\alpha_c(z,t)=\sigma_c N_c(z,t)$ accounts for the FCA losses, where $N_c(z,t)$ is the free-carrier density and $\sigma_c=1.45\times10^{-21}$ m$^2$ for silicon \cite{Lin:07}. $\mu(z,t)=2k_c(z,t)\omega_{p}/(\sigma_c c)$ is the FCD where $c$ is the speed of light and $k_c(z,t)=(8.8\times 10^{-28} N_c(z,t) + 1.35 \times 10^{-22} N_c(z,t)^{0.8})/N_c(z,t)$ is the free carrier index \cite{Lin:07}. 

The dynamics of the carriers density $N_c(z,t)$ is described by Eq.~(\ref{FC}), which contains three terms accounting respectively for the generation of free carriers through the two-photon absorption and the avalanche ionization, as well as the recombination rate. In this equation, $\hbar$ is the reduced planck constant, $\sigma_{AI}=\mu_0q_e^2c/(m_e\tau_0\omega_p^2)$ is the avalanche ionization cross section \cite{Skupin:06}, where $\mu_0$ is the vacuum permeability, $q_e$ is the elementary charge, $m_e$ is the electron mass at rest and $\tau_0=20$ fs is the electron collision time. $U_i=1.12$ eV is the silicon bandgap energy, $a$ is the mode area and $\tau_c=1$ ns is the carrier lifetime. For the conditions considered in this study, the free carriers generated through the avalanche ionization effect are one order of magnitude lower than those generated by two-photon absorption. The small effect of the avalanche ionization is due to the long pulse duration ($\approx$ 200\,fs) and low energy pulse ($\approx$ 5\,pJ) at the waveguide input. However, for input pulses of the order of 70\,fs with an input energy of $\approx100$\,pJ, the avalanche ionization would be the leading mechanism in the generation of free carriers. Finally, all the waveguide parameters in the Eqs.~(\ref{EqGNLS}-\ref{FC}) are evaluated at each propagation step as the waveguide width is $z$ dependent.

\section{Results and discussions}

\begin{figure}[t!]
\centering\includegraphics[width=\linewidth]{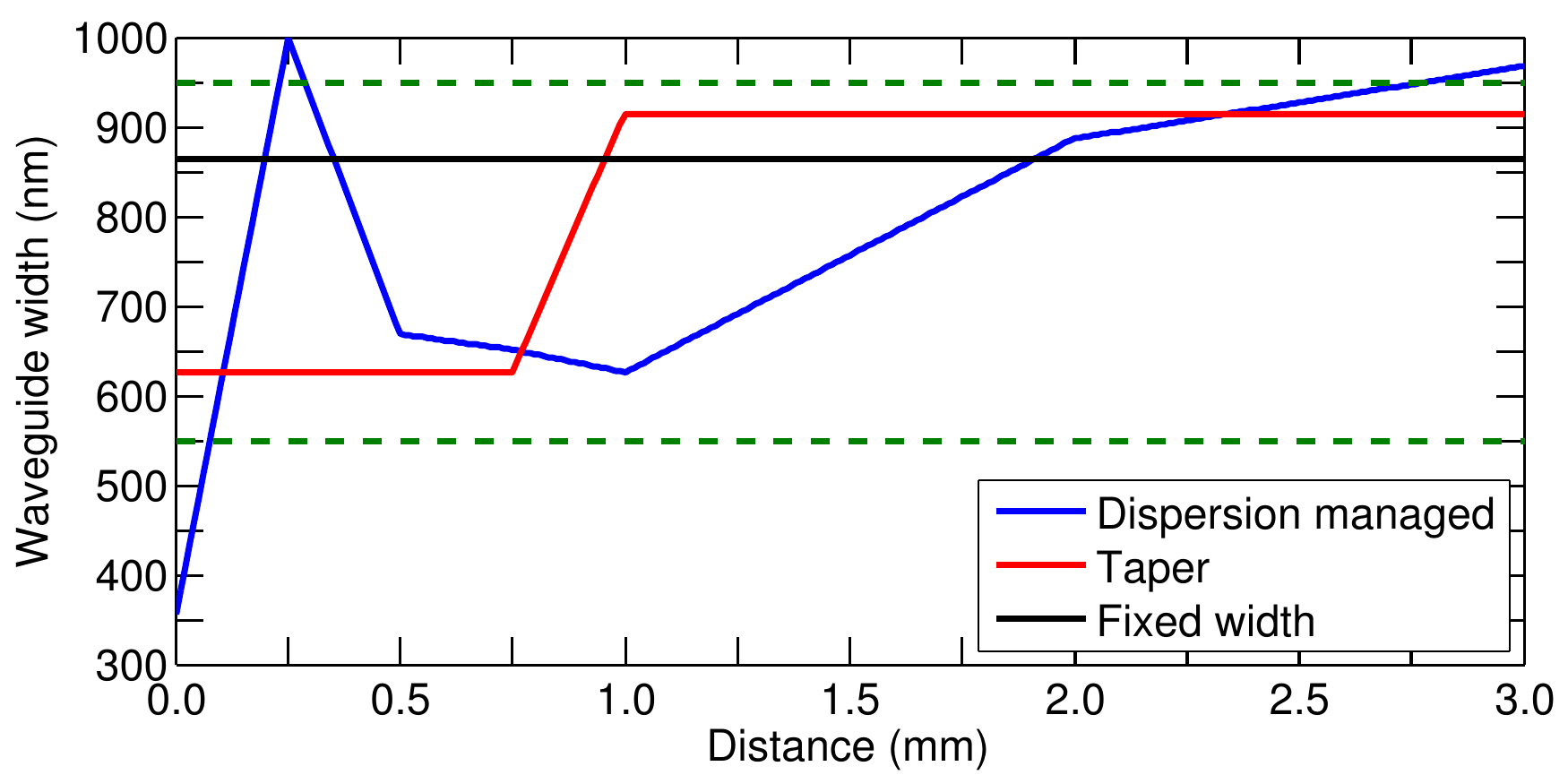}
\caption{(Color online) Waveguide width profile for the dispersion managed (blue curve), tapered (red curve) and fixed width (black curve) waveguides. All these profiles have been optimized with the RMHC method. The two green dashed lines refer to the waveguide width corresponding to one of the two zero-dispersion wavelengths located exactly at the 2\,$\mu$m pump wavelength (see also in Fig\,\ref{GVD}). The anomalous (normal) dispersion regime is thus located between (outside) these two dashed lines.}
\label{Width}
\end{figure}

\begin{figure}[b!]
\centering\includegraphics[width=\linewidth]{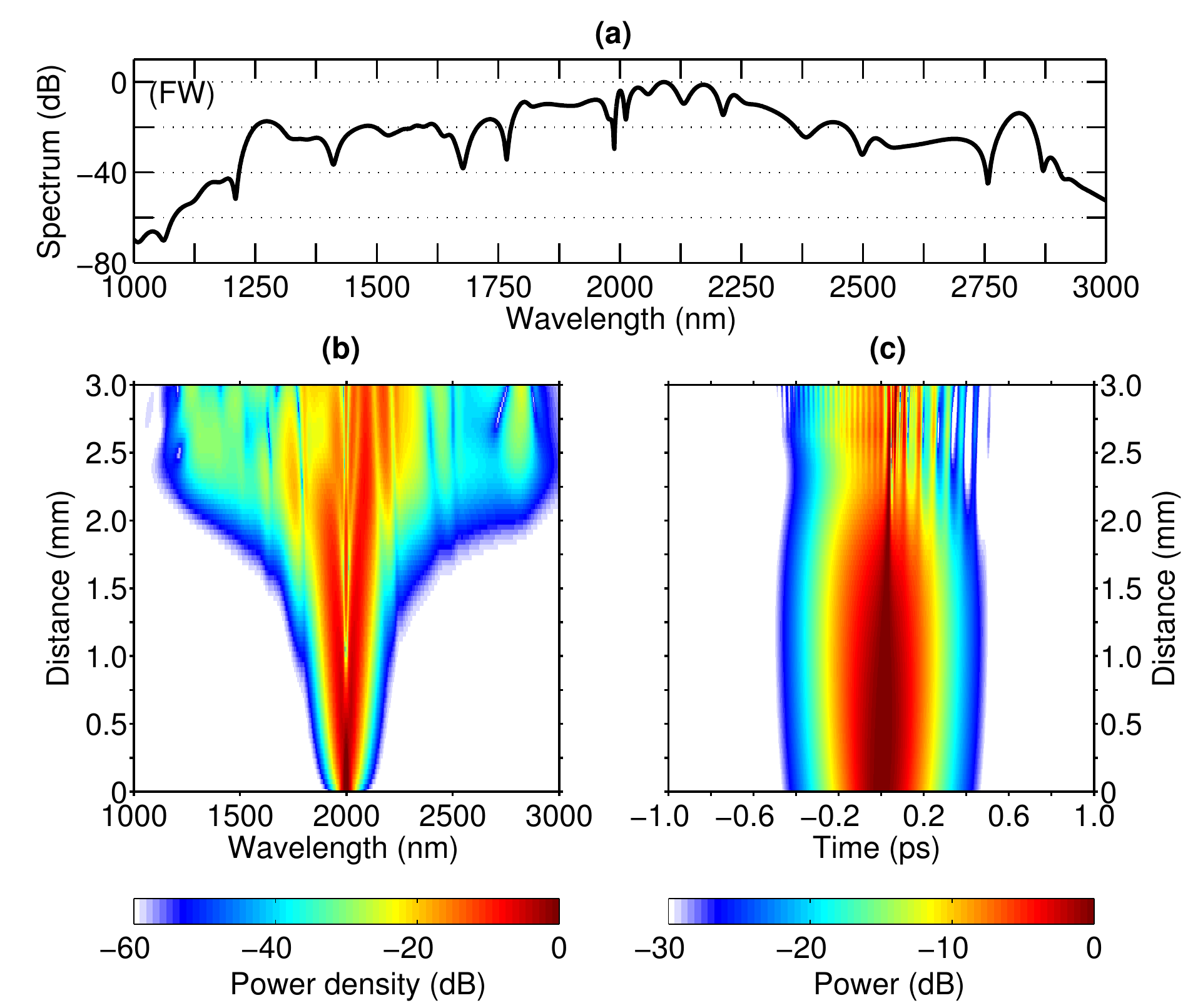}
\caption{(Color online) Simulated normalized spectrum for a 32~W, 200 fs input pump wave at 2000 nm, at the output of the 3 mm-long fixed width waveguide (a). Density plots (in pseudocolors) of the simulated spectral (b) and temporal (c) evolutions with the position in the waveguide. In these two figures, the amplitudes are normalized with respect to the maximum input spectral density.  FW: fixed width.}
\label{FW}
\end{figure}

The Fig.~\ref{Width} shows the width profiles $w(z)$ for the tapered and the DM waveguides that optimize the fitness function~(\ref{fitness}) for a 32~W peak power, 200~fs input pump pulse at 2000 nm. In order to show the potential benefit of these structures over fixed width waveguides, all the structures have been computed with the same RMHC algorithm to find their optimal profile. It can be seen that for the FW and the tapered waveguide the dispersion remains anomalous at the pump wavelength. It is also the case for the DM waveguide except for one brief excursion in the normal dispersion region at the beginning of the propagation, as well as at the waveguide end. 

\begin{figure}[b!]
\centering\includegraphics[width=\linewidth]{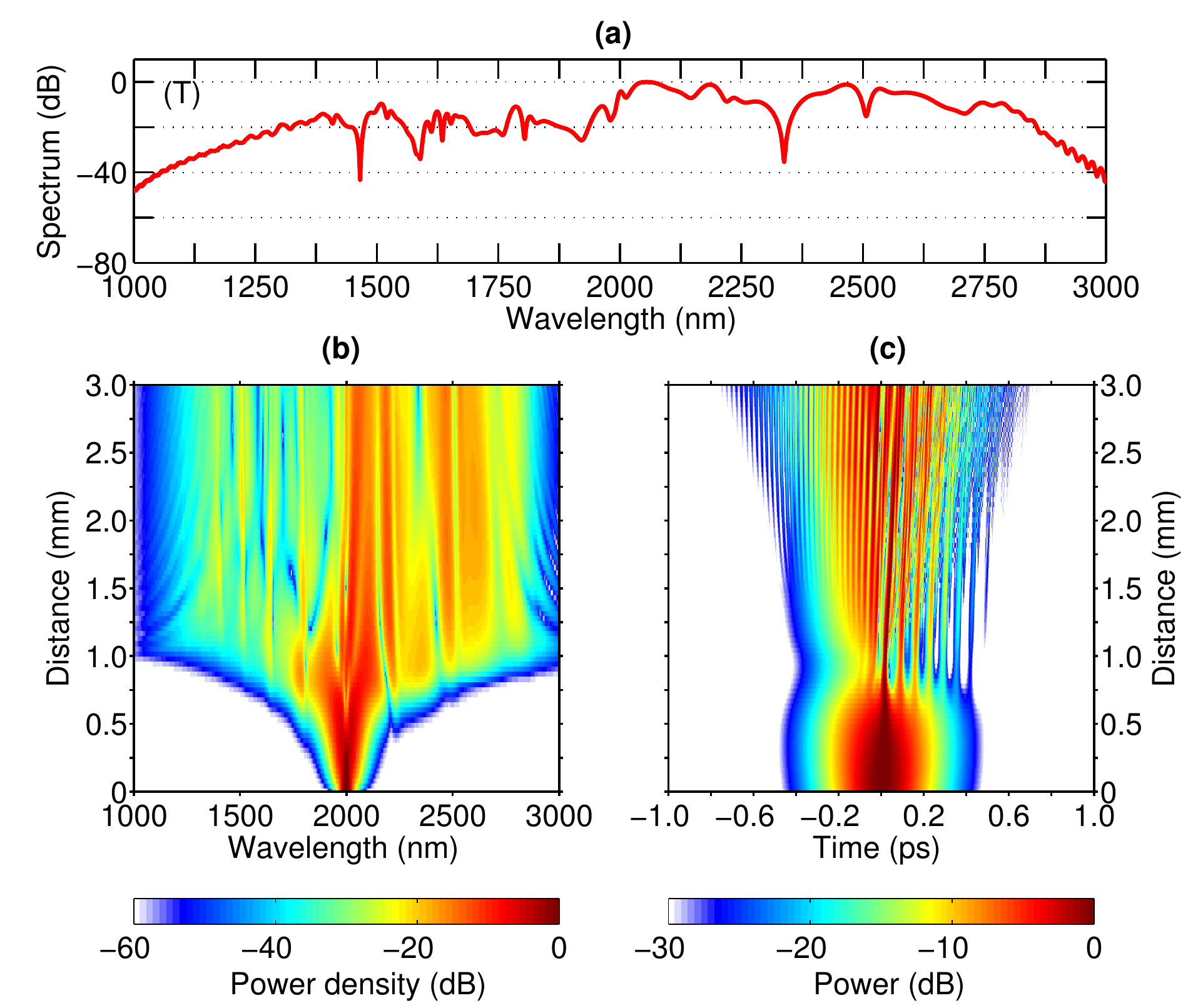}
\caption{(Color online) Supercontinuum generation in the tapered (T) waveguide characterized by the dispersion map displayed in Fig.\,\ref{Width}. Same initial conditions as in Fig.\,\ref{FW}.}
\label{Taper}
\end{figure}

\begin{figure}[t!]
\centering\includegraphics[width=\linewidth]{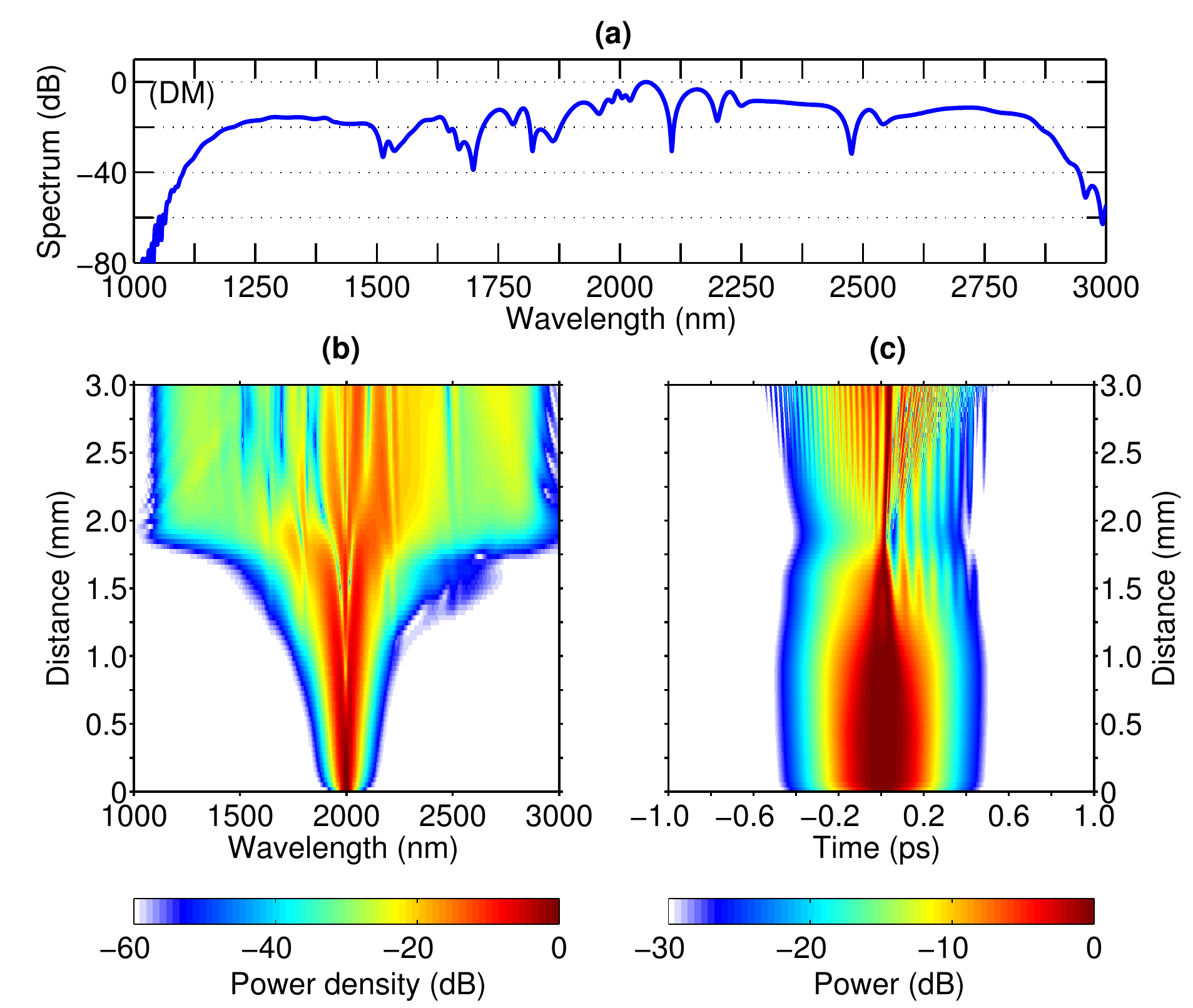}
\caption{(Color online) Supercontinuum generation in the dispersion managed (DM) waveguide characterized by the dispersion map displayed in Fig.\,\ref{Width}. Same initial conditions as in figure\,\ref{FW}.}
\label{DM}
\end{figure}

The computed output spectra generated by propagation in the optimal fixed width waveguide ($f=0.15)$, in the taper structure ($f=0.20)$ and in the DM structure ($f=0.26$) are shown in Fig.~\ref{FW}(a), Fig.~\ref{Taper}(a) and Fig.~\ref{DM}(a), respectively. If we consider the bandwidth starting from the shortest wavelength up to the longest one corresponding to the spectral components crossing the -20 dB level, the spectrum starts at 1250\,nm and ends at 2850 nm for the optimal FW waveguide [see in Fig.\,\ref{FW}(a)]. However, these limits are mainly due to the emergence of two DWs at shorter and longer wavelengths. Between these two DWs, there are large wavelength ranges where the normalized power density decreases significantly below -20 dB. Consequently, if the two DWs are not considered, the spectrum (at -20\,dB) extends only from 1700\,nm to 2360\,nm. In the tapered waveguide, the spectrum spans [1300 -- 2860\,nm] at the -20\,dB limit [see in Fig.~\ref{Taper}(a)] and shows less amplitude variations, resulting in a higher $f$ value. Decreasing dispersion configuration has already been identified as a mean to improve the SC spectral width \cite{Abrardi:07,Stark:12,Hu:15}. Note that contrary to SC generation in engineered taper fibers, the dispersion in the first section is not normal. 

Even though tapered structures already improve the broadening mechanism, other dispersion maps could potentially give rise to further improvement of the SC spectra. As seen in the figures, the SC generated in the optimally designed DM waveguide, is not only slightly broader than the SC generated in the tapered structure [1200 -- 2870\,nm], but is also flatter, in particular in the range [1.1--1.5 $\mu$m]. The spectrum generated in the DM waveguide is indeed more symmetric with respect to the pump wavelength, showing the potential of the DM designs over simpler taper designs. The root-mean square spectral width of the DM spectrum is 163\,radTHz, a value larger than the spectral width at the output of the taper structure (130\,radThz) or of the FW waveguide (109\,radTHz).    

In order to gain further insights on the overall dynamics of the supercontinuum generation, the spectral and the temporal evolutions of the pulse along the 3\,mm propagation are shown for each waveguide designs in Fig.~\ref{FW}(b,c)~Fig.~\ref{Taper}(b,c)~Fig.~\ref{DM}(b,c). It can be seen in these figures that, at the location where the pulse is the most temporally compressed (i.e. for instance at a distance of 1.8\,mm for the DW waveguide) the waveguide width is 865\,nm for the taper and the DM waveguides. This value is identical to the optimal width of the FW structure. We can thus infer that this width maximizes the generation of two DWs far from the 2\,$\mu$m pump wavelength. Taking the local dispersion value and the pulse amplitude and width into account at this particular point, the order of the soliton is 3.6 in the FW waveguide, while it is equal to 1.4 in the DM waveguide and to 1.5 in the taper waveguide.

The propagation in the DM waveguide can be separated into three distinct sections: section 1 [0-1\,mm], section 2 [1-2\,mm] and section 3 [2-3\,mm]. In the first section, the fast tapered sections aim at shaping the input pulse to optimize the spectral broadening in sections 2 and 3. The spectral width slightly increases before entering the section 2, but not as much as before the beginning of the taper in the tapered waveguide. The second section is a standard taper in which the dispersion at the pump wavelength is scanned across the anomalous dispersion region, similarly to the optimized tapered waveguide (see curves in Fig.~\ref{Width}). It is in this section that the initial pulse is strongly temporally compressed and that the DWs at shorter and longer wavelengths are efficiently excited. In the tapered waveguide, it can be seen that at the location of the maximum temporal compression, delayed satellite pulses are clearly visible. These satellite pulses are weaker in the DM waveguide and this difference might be the result of the propagation in the first dispersion varying section and leads to a smoother spectral profile. Finally, the last section of the DM dispersion map is a second tapered section with a weaker gradient in which the dispersion profile evolves toward a flat curve with a small amplitude across the whole spectrum. The variation of the dispersion encountered in this last section enables the phase matching condition for DW generation to sweep through the 1200-1500\,nm and 2500-2800\,nm regions, leading to an increase of the spectral density in these wavelength ranges. At the end of the propagation, the dispersion on the whole spectrum is normal and close to the blue curve shown in Fig.\,\ref{GVD} (1000\,nm width waveguide). 

In our study, we have considered a 32\,W, 200\,fs input pulse at 2\,$\mu$m. It is nevertheless expected that the optimized dispersion map depends on these parameters. The simulations with the same dispersion map as in Fig.\,\ref{Width} however reveal that the generated SC in the DM waveguide is not very sensitive to the input power or to the input pulse wavelength. A variation of $\pm$40\,$\%$ of the input power gives similar results as in Fig.\,\ref{DM} with only a 1.5$\%$ decrease of the spectral width at -40\,dB. Similarly, the input wavelength can be tuned in the range 1900--2200\,nm without any detrimental consequences. Note that beyond 2.2\,$\mu$m, the model should be revised as the three-photon absorption becomes the dominant nonlinear loss mechanism. On the contrary, the input pulse duration is a more critical parameter. The optimized design is nevertheless robust against a variation of $\pm$~15~fs of the pulse duration. Finally, we have numerically study the robustness of the SC generation in the DM waveguide against fabrication tolerances. To this end, we have considered a variation of the waveguide width over the whole profile of $-20$\,nm and $+20$\,nm with respect to the optimal design, a tolerance easily achievable with state-of-the art fabrication processes. Randomly distributed width variations between $-20$ and $+20$\,nm at each optimization points have also been considered. Interestingly for applications, no detrimental consequences have been observed to the generated supercontinuum as it remains always broader and flatter than those obtained with the optimal FW and tapered structures.

In some important applications large SC bandwidth but also high coherence are needed. The degree of coherence of the SC generated in the different structures have thus been studied through the degree of first order coherence  \citep{Leo:15}
\begin{equation}
\left|g_{12}^{(1)}(\lambda)\right|=\left|\frac{\langle A_1^\star(\lambda)A_2(\lambda)\rangle}{\sqrt{\langle|A_1(\lambda)|^2\rangle\langle|A_2(\lambda)|^2\rangle}}\right|,
\end{equation}
 where the angle brackets denote ensemble averages over independent realizations of spectra pairs $A_{1,2}(\lambda)$ obtained from 200 separate simulations with different realization of input noise \cite{Dudley:06}, $A_{noise}(\omega)=\sqrt{\hbar\omega/ T_{span}}\exp{\left[i2\pi\varphi(\omega)\right]}$, corresponding to adding one photon per spectral discretization of the electric field with random phase, into the initial condition. $T_{span}$ is the temporal window used for the simulations (typically 15\,ps). $\varphi(\omega)$ is a random phase between 0 and $2\pi$ with a Gaussian distribution. We also computed the mean coherence $\overline{g_{12}}$, averaged over the bandwidth $[\omega_1-\omega_2]$, where $\omega_{1,2}$ are the shortest and the largest angular frequency that crosses the -20\,dB level. Note that a coherence criteria could be added to the fitness function $f$, by adding a third term in Eq.~(\ref{fitness}) equivalent to the average spectral coherence. Optimizing this new fitness function would also result in a maximization of the overall coherence. However the computing of $g_{12}^{(1)}$ is very time consuming as it requires a sufficiently large set of independent simulations. 

  \begin{figure}[t!]
\centering\includegraphics[width=\linewidth]{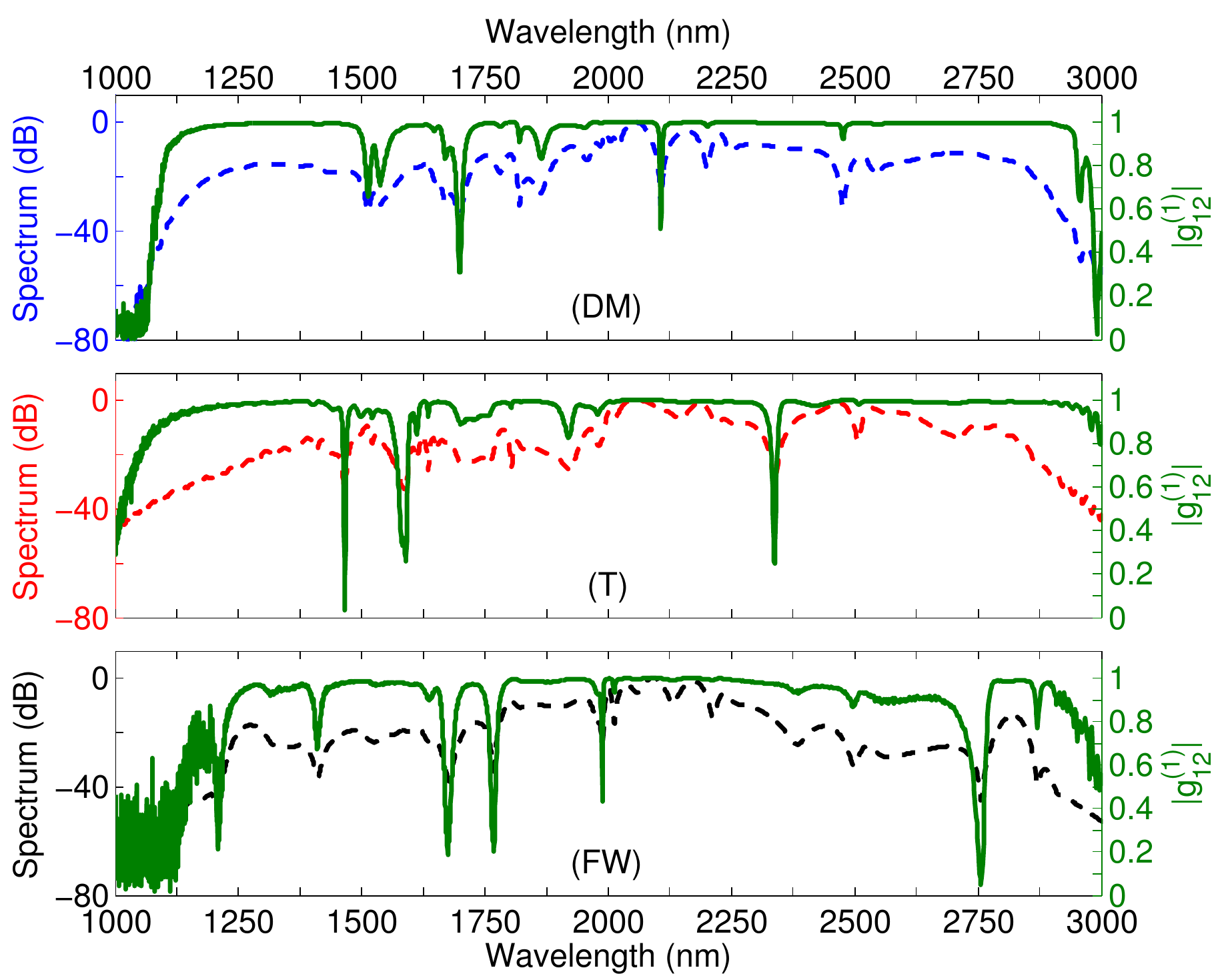}
\caption{(Color online) Simulated degree of first order coherence $\left|g^{(1)}_{12}\right|$ for each considered structures (green curves). The output spectrum are also plotted as dashed curves. DM: dispersion managed, T: tapered and FW: fixed width.}
\label{Coherence}
\end{figure}

The results reported in Fig.\,\ref{Coherence} show that a very high coherence is maintained over the whole SC spectrum generated in the DM waveguide, which is confirmed by the high value of the averaged coherence, $\overline{g_{12}}=0.98$. The coherence is slightly lower for the tapered structure, $\overline{g_{12}}=0.97$.  A significant decrease of the coherence in part of the spectrum is however visible for the FW waveguide, which is also highlighted by a lower value of the averaged coherence, $\overline{g_{12}}=0.91$. The lower coherence encountered for the FW waveguide can readily be explain from the fact that the input pulse corresponds to a $N=12$ soliton. As the pump wavelength is at 2\,$\mu$m, the two-photon absorption is not strong enough to sufficiently quench the modulation instability as in\,\citep{Leo:15} where SC generation in silicon FW waveguide at 1550\,nm was considered. On the contrary, in the taper and DM waveguides, the soliton order remains below 8 during the whole propagation which ensures a high coherence\,\citep{Dudley:06} despite the relatively long (200\,fs) initial pulse considered.

\section{Conclusion}

To summarize, we studied SC generation in silcon nanophotonic taper waveguide as well as in more elaborated dispersion managed waveguide, in which the group-velocity dispersion is optimally adjusted during the propagation to improve the generation of  supercontinuum. In semiconductor on insulator nanophotonic waveguides, the dispersion properties are strongly dependent on the waveguide width. The proposed tapered and DM structures thus show a longitudinal variation of the waveguide width to manage the dispersion encountered by the pulse during its dramatic spectral broadening. Hence, the considered DM waveguides are more complex but also easier to fabricate compared with glass based taper planar waveguide previously considered\,\citep{Hu:15} for SC generation. The optimization of the design of the structure resorts to a RMHC genetic algorithm to generate the broadest and the flattest SC achievable regarding the input pulse and the design constraints. An ultra broadband supercontinuum spanning 1100 nm - 2950 nm at a spectrum level of -40 dB has been demonstrated in silicon on insulator DW waveguides. It is shown that the generation of dispersive waves is the leading mechanism enabling the generation of new frequencies far from the pump wavelength. However, thanks to the versatility of the structure, the SC generated in the DM waveguide is slightly broader and flatter compared to the results in the optimal tapered structure and largely better than the commonly used fixed width waveguides. It is also demonstrated that the SC shows excellent coherence properties in DM and taper structures, better that the coherence obtained for the simplest fixed width waveguide. This demonstrates the potentialities of these two structures for on chip SC generation starting from rather long pulses (200\,fs) and very low energies (<10\,pJ) as they are not more complicated to fabricate than the well known dispersion engineered fixed width waveguides. This approach could also been proven to be useful with other integrated platforms such as chalcogenides, InGaP, SiN, etc.; for generating spectra with desired particular features or for the generation of frequency combs in resonators as long as their dimensions allow for the variation of the waveguide width.      

\section*{Funding}

This work is supported by the Belgian Science Policy Office (BELSPO) Interuniversity Attraction Pole (IAP) project Photonics@be and by the Fonds de la Recherche Fondamentale Collective, Grant No. PDR.T.1084.15. 

\section*{Acknowledgment}

The authors thank G. Steinmeyer and L. Bergé for the fruitful discussions on the effect of free carriers generation through avalanche ionization.





\end{document}